\documentclass[
prl,
twocolumn,
superscriptaddress,
floatfix
]{revtex4-2}

\usepackage{amsmath,latexsym,amsfonts,amssymb}
\usepackage{natbib}
\usepackage{epsfig}
\usepackage{graphicx}
\usepackage{inputenc}
\usepackage{setspace}
\usepackage{multirow}
\usepackage{array}
\usepackage{tabularx}
\usepackage{bm}
\usepackage{color}
\usepackage{xcolor}
\usepackage{gensymb}
\usepackage[left]{lineno}
\usepackage{ulem}
\usepackage{siunitx}
\usepackage{booktabs}
\usepackage{subcaption}

\newcommand{\LNO}{La$_3$Ni$_2$O$_7$}
\newcommand{\LSNO}{La$_2$SmNi$_2$O$_7$}

\begin{document}

\title{Pressure-Driven Structural Transitions without a Displacive Charge-Density Wave in \LSNO}

\author{J. Huang}
\affiliation{Universit\'e Paris-Saclay, CNRS, Laboratoire de Physique des Solides, 91405, Orsay, France.}
\affiliation{Synchrotron SOLEIL, L\'\ Orme des Merisiers, Saint Aubin BP 48, 91192, Gif-sur-Yvette, France}

\author{Sitaram Ramakrishnan}
\affiliation{CNRS, Universit\'e Grenoble Alpes, Institut N\'eel, F-38000 Grenoble, France}

\author{P. Rodi\`ere}
\affiliation{CNRS, Universit\'e Grenoble Alpes, Institut N\'eel, F-38000 Grenoble, France}

\author{P. Toulemonde}
\affiliation{CNRS, Universit\'e Grenoble Alpes, Institut N\'eel, F-38000 Grenoble, France}

\author{Z. Rahmany}
\affiliation{Universit\'e Paris-Saclay, CNRS, Laboratoire de Physique des Solides, 91405, Orsay, France.}

\author{V. Bal\'edent}
\affiliation{Universit\'e Paris-Saclay, CNRS, Laboratoire de Physique des Solides, 91405, Orsay, France.}
\affiliation{Institut universitaire de France (IUF)}

\author{B. Vignolle}
\affiliation{Institut de Chimie de la mati\`ere condens\'ee de Bordeaux, Universit\'e Bordeaux, France}

\author{Sourav Marik}
\affiliation{Institut de Chimie de la mati\`ere condens\'ee de Bordeaux, Universit\'e Bordeaux, France}
\affiliation{Department of Physics and Materials Science, Thapar Institute of Engineering and Technology, Patiala 147004, India}

\author{P. Fertey}
\affiliation{Synchrotron SOLEIL, L\'\ Orme des Merisiers, Saint Aubin BP 48, 91192, Gif-sur-Yvette, France}

\author{P. Foury-Leylekian}
\affiliation{Universit\'e Paris-Saclay, CNRS, Laboratoire de Physique des Solides, 91405, Orsay, France.}
\email [Corresponding author: ] {pascale.foury@universite-paris-saclay.fr}

\begin{abstract}
We investigated the structural properties of bilayer nickelate \LSNO{} as a function of pressure and temperature. At ambient conditions, we show that the material crystallizes as a monoclinic superstructure distinct from the one previously reported and close to the pseudo-orthorhombic structure of pristine \LNO{}.  No signatures of satellite reflections associated with charge density wave (CDW) ordering are detected at low temperature. Upon compression, a sequence of pressure-induced structural transitions from monoclinic to orthorhombic ($15$ $GPa$) and then tetragonal ($21$ $GPa$) symmetry is observed. Within the superconducting dome, the quality of the X-ray diffraction data enables structural refinements enabling theoretical models to understand the emergence of superconductivity.
\end{abstract}

\maketitle

The recent discovery of a superconducting phase at a temperature as high as $80$ $K$ under pressures above $14$ $GPa$ in the bilayer Ruddlesden-Popper nickelate \LNO{} \cite{Sun, MWang, Hangsupravraie} has opened a new frontier in the field of high-temperature superconductor (SC) outside the cuprate \cite{Keimer} and pnictide families \cite{Ishida, Si}. This finding raises fundamental questions regarding the role of Ni 3d electronic correlations, interlayer coupling, and spin fluctuations in mediating Cooper pairing in a system where both $dx^2$-$y^2$ and $dz^2$ Ni orbitals are expected to be active near the Fermi level. 

It is evidenced that the crystal structure of \LNO{} plays a central role in the SC properties. At ambient conditions, the compound adopts an orthorhombic structure (space group $Amam$) characterized by a significant buckling of the NiO$_6$ octahedra, resulting in Ni–O–Ni bond angles along $c$ substantially deviating from 180$\degree$, which reduces the inter-plane hopping and interlayer superexchange coupling favorable to SC \cite{ChenLu}. Under applied pressure at room temperature, a first work showed that the system undergoes a structural phase transition toward a higher symmetry structure corresponding to a $Fmmm$ space group in which the Ni–O–Ni angles along $c$ approach linearity \cite{Sun}. Further investigations identified an orthorhombic-to-tetragonal transition associated with the $I4/mmm$ space group at low temperature and under pressure \cite{WangL2024} leading to a strictly linear arrangement of Ni-O-Ni. This last structure is widely believed to be connected with the emergence of superconductivity, as the SC onset broadly coincides with the pressure range over which the structural transition occurs. Nevertheless, recent experimental evidence has demonstrated that superconductivity can persist in both orthorhombic and tetragonal structural phases, suggesting that while the straightening of the Ni–O–Ni network is favorable for pairing, it may not constitute a strict prerequisite for superconductivity in this system \cite{Shiortho}. However, the question of the symmetry of the space group favorable to superconductivity remains an open question.

A key challenge lies in stabilizing the superconducting phase at lower external pressure. Chemical substitution on the La site notably, offers a route to mimic applied pressure via chemical pressure without disruption of the Ni network. Beyond simply decreasing the SC critical pressure and potentially raising T$_c$, chemical pressure constitutes a valuable tool to map out the generalized phase diagram of \LNO. In close analogy with cuprates and iron-based superconductors \cite{Si, Ishida, Keimer}, the nickelates host interesting competing states such as spin-density wave (SDW) and charge-density wave (CDW) phases below an ordering temperature T$_{DW}$ close to $150$ $K$ \cite{Liu2022,Xu,GWangDW2024,LiuZ2024,Khasanov2025,Kakoi,YiDFT}, that may be precursor of superconductivity or compete with it. Chemical substitution allows tuning across these transitions but also the structural transitions associated to the Ni-O-Ni angle linearity or to the tetragonal symmetry. Accurately establishing such a generalized phase diagram is crucial to disentangle the role of lattice symmetry and charge or spin orders on Cooper pairing, and to identify the key ingredients driving high-T$_c$ superconductivity in nickelates.

A major bottleneck in the study of \LNO{} is the inherent difficulty of its synthesis. The stabilization of the phase requires stringent control of oxygen pressure and growth conditions \cite{Sun}. Bulk single crystals present typically intergrowths \cite{Puphal, Wang, Chen}, oxygen vacancies, and phase inhomogeneities that severely compromise reproducibility of the SC properties. Achieving high-quality samples thus appears as a prerequisite for any reliable mapping of the phase diagram and for unambiguous assessment of the intrinsic SC properties of the system. Remarkably, chemical substitution on the rare earth site has been shown to play a beneficial role not only in tuning the electronic structure, but also in improving phase purity and crystalline order and suppress competing secondary phases \cite{NWang, ZDong}. 


Among the various chemical substitution strategies, replacement of La by Sm stands out as particularly promising, as theoretical calculations predicted a doubling of T$_c$ upon full substitution \cite{Pan}. However, due to chemical constraints, synthesis has been limited to a maximum Sm content of 1.5 per formula unit \cite{Li, Zhong2025}. In La$_2$SmNi$_2$O$_{7-\delta}$ single crystals synthesized at ambient pressure, Li et al. have evidenced a monoclinic $P2_1/m$ structure ($a = 5.45$ \text{\AA}, $b= 5.36$ \text{\AA}, $c= 10.50$ \text{\AA} $\beta$ $\approx$ 105$\degree$ see Fig. S4 of Supplemental Information (SI)) and demonstrated metallic properties at ambient conditions. Interestingly, no resistivity anomaly characteristic of a possible density wave instability has been observed unlike in pristine \LNO \cite{Liu2022}. For pressurized \LSNO{}, a tetragonal structure is detected above 20 GPa at 300 K as well as a a superconducting dome starting from 14.8 GPa with a maximum critical temperature at 21.6 GPa (SC onset at $T_c$=92 K and zero resistance at 73 K) \cite{Li}. The study thus seems to show that both monoclinic and tetragonal structures can support superconductivity. Remarkably, further increasing the Sm content to 1.5, does not significantly improve $T_c$ \cite{Zhong2025}.  

Until now, most structural investigations under pressure on \LNO and parent compounds such as \LSNO{} have only been performed on powder or at room temperature. Single-crystal x-ray diffraction experiment under simultaneous low-temperature and high-pressure conditions, where superconductivity emerges, remains largely unexplored. Yet an accurate determination of the crystal structure in the SC state is essential to unambiguously establish the symmetry, quantify the interlayer distances relevant in the coupling of NiO bilayers through the apical oxygen, resolve the origin of the orthorhombic-to-tetragonal transition and its exact role in the stabilization of the SC phase. 

In this paper, we present a detailed synchrotron x-ray diffraction study as a function of pressure performed at the CRISTAL beamline of synchrotron SOLEIL, on both \LSNO{} powder and single crystals at ambient and low temperature. We report a deviation from the previously published ambient conditions structure related to oxygen displacements. We also show that if a displacive CDW exists, the atomic displacement must be less than few thousandths of an Angstr$\Ddot{o}$m. Finally, by tracking the structural evolution under pressure and low temperature, we provide accurate structural parameters across the superconducting phase and its precursor state. These results shed new light on the interplay between lattice, electronic and magnetic degrees of freedom.

Single crystals were grown under ambient-pressure conditions similar to reference \cite{Li}. High-purity La$_2$O$_3$ (99.99\%) and Sm$_2$O$_3$ (99.99\%) powders were preheated at 800 \degree C for 6 h prior to use. Stoichiometric amounts of the rare-earth oxides and NiO powders (99.99\%) were mixed and ground well before being loaded into an Al$_2$O$_3$ crucible. The mixture was then combined with anhydrous K$_2$C$O_3$ flux at a mass ratio of 1:15. The crucible was covered with a lid, and the loading of the mixture of the sample and anhydrous K$_2$C$O_3$ was carried out inside a glove box. Crystal growth was achieved by heating the sample at 1050 $\degree$ C for 72 hours, followed by slow cooling at a rate of 1 $\degree$ C/h. Laboratory x-ray measurements has been conducted to check the crystallographic phase and sample quality. The chemical composition of the products was analyzed by energy dispersive x-ray spectrometer equipped with a scanning electron microscope (SEM/EDX). We took an average of chemical compositions of 10 points for two single crystals and obtained a stoichiometry for La:Ni of 0.96 $\pm{0.03}$ and La:Sm 2.1 $\pm{0.05}$ similar to \cite{Li}. All measurements reported in this work were performed on samples originating from the same synthesis batch, ensuring a consistent comparison between different experimental probes.

X-ray diffraction experiments were carried at the SOLEIL synchrotron on the CRISTAL beamline. Single crystal diffraction at ambient pressure and low temperatures were conducted with a Newport 4-circle diffractometer in a Kappa configuration equipped with a Rigaku Oxford Diffraction Atlas CCD detector and a He blower, at a wavelength around $0.67$ $\AA$. The measurement under pressure at high and low temperature were conducted with a Newport six-circle diffractometer equipped with a 2D GaAs hybrid pixels detector from the XSpectrum company, at a wavelength of $0.4165$ $\AA$. Samples with a volume of approximately 30*30*25 $\mu m^3$ were mounted in a Diamond Anvil Cell (DAC) using He gas as the pressure transmitting medium and a stainless steel gasket. The DAC was placed in a He flow cryostat allowing a rotation around the vertical axis. The pressure was measured using the ruby fluorescence technique. The measurements were performed as a function of pressure ($0–25$ $GPa$) at $10$ $K$, yielding a total of approximately 1500 Bragg reflections at each (P, T) point. The powder measurements under pressure were conducted only at room temperature. A tiny part of the washed batch composed of \LSNO{} single crystals, was carefully ground into powder and loaded into the hole of a rhenium gasket mounted on the diamond of the DAC. This loaded powder was preliminary measured by x-ray diffraction using a laboratory microfocused x-ray source before the SOLEIL experiment to check for the quality of the pattern. The lattice parameter of few polycrystalline grains of cubic NaCl were used as an internal pressure calibrant, in addition to the ruby fluorescence. Structural refinements were performed by the JANA2020 program \cite{PetricekVaclav2014}.


Single-crystal x-ray diffraction measurements at $300$ $K$ and $15$ $K$ repeated on 3 different samples, reveal systematically a crystal structure different than that previously reported \cite{Li}. A new feature appears in the reconstruction of the $(0kl)$ reciprocal plane at $300$ $K$ as shown in Fig. 1 (see also Fig. S1 of SI for measurements at $15$ $K$). It evidences the presence of weak superlattice reflections having the experimental resolution and located at $l/2$ $c^*$. These superlattice reflections imply a doubling of the $c$ unit cell parameter. Their intensity being 10$^{-2}$ order of magnitude less than the principal Bragg reflections, they might have been missed in previous measurements performed with laboratory x-ray sources or on powder. Starting from the monoclinic structure published by Li et al, there are two possible subgroups of k-index 2 : $P2_1/m$ and $P2_1/c$. By studying the $(h0l)$ reciprocal plane indexed in the monoclinic c-doubled lattice (see Fig. S2 of SI), we observe the systematic absence of Bragg reflections with $l$ odd, characteristic of the extinction rules of a $c$ glide mirror. This doubled $P2_1/c$ (b unique) monoclinic cell can also be described in a $P2_1/n$ (a unique) pseudo-orthorhombic cell close to the one of the $Amam$ structure determined for \LNO{} \cite{Sun}($\overrightarrow{c_{{PO}}} = 2\overrightarrow{c}+\overrightarrow{a}$). Fig S4 of the supplemental information shows the various conversions of unit cell parameters. Least-squares refinement in the $P2_1/n$ space group converged to $R_{\mathrm{obs}} = 7.61\%$ at 15~K and
$R_{\mathrm{obs}} = 6.10\%$ at 300~K.
Details of the refinements are given in Tables S1 and S2. From these results, one can understand that the doubling of the unit cell along $c$ compared to the initial monoclinic structure of Li et al \cite{Li,Zhong2025} does not arise from the La/Sm and Ni atomic displacements or occupancies rates because these atoms are located on the $n$ glide mirror and are related by a $c/2$ translation symmetry. The cell doubling is only due to the O4 and O5 atoms, out of the $n$ mirror. These displacements, detailed in the SI, produce a distortion of the NiO6 octahedra that alternates in direction from one Ni–O bilayer to the next, giving rise to an antiferrodistortive-like structural motif. As seen in Fig. \ref{Fig_2}, these oxygens belong to the Ni-O planes and are not involved in the Ni-O-Ni angle along $c$. The refinement in $P2_1/n$ finally leads to similar occupancy in the various La sites than on Li et al model with a weak Sm occupancy in the inner layers (11\% vs 7\%). This is also in perfect agreement with recent DFT calculations \cite{Zhao}. Finally, we studied the Ni–Ni distance along $c$ which is related to the coupling of NiO layers, a central point for the emergence of SC. We observe that this distance weakly contracts by 0.4\% upon cooling from room temperature to $15$ $K$, decreasing from 3.949(7) \text{\AA}\ at $300$ $K$ to 3.933 (5) \text{\AA}\ at $15$ $K$. This favors the hybridation between Ni orbitals and increases the coupling within the bilayer which is already strong in RP nickelate SC.

We investigated as well, the possibility of a structural modulation associated with the CDW phase at low temperature, even in the absence of clear evidence of a density-wave transition in the resistivity measurements of Li et al \cite{Li}. To this end, we searched for the presence of satellite reflections in the reconstructions of the (a*,b*) reciprocal planes at $15$ $K$, below the critical temperature expected for the possible CDW transition ($\approx$ $150$ $K$) which are absent at $300$ $K$. Two scenarios can be envisioned. In the first, the CDW originates from Fermi surface nesting, which would give rise to a propagation wave vector (commensurate or not) close to $q_1\approx$(0 $\frac{1}{2}$ 0) or $q_2\approx$($\frac{1}{2}$ $\frac{1}{2}$ 0) according to band structure calculations and recent ARPES measurements on pristine \LNO{} \cite{ARPES}. In the second, the CDW is driven by an exchange-striction effect associated with a SDW, as in Chromium \cite{Tsunoda}, in which case the propagation wave vector of the structural modulation would be at 2$q_1$ or 2$q_2$. Despite a sensitivity reaching 5$*$10$^{-5}$ times the intensity of the standard Bragg reflections, no additional reflections were detected at $15$ $K$ (see Fig S3 of SI). As the intensity expected for a CDW is proportional to the square of the atomic displacement \cite{Pouget}, the upper bound on any possible CDW atomic modulation must be less than few thousandths of 1 \AA. 

\begin{figure}[htbp]
\centering
\includegraphics[width=0.9\linewidth]{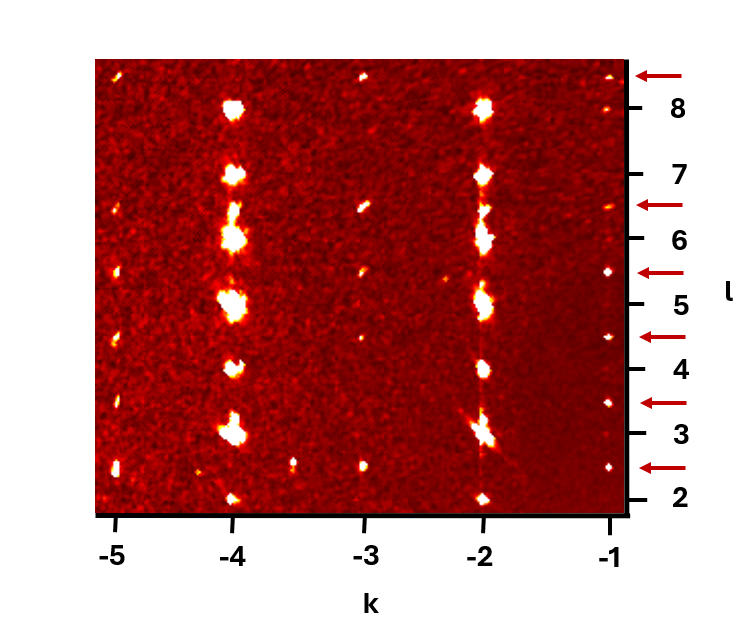}
\caption{\footnotesize(Color online) Reconstruction of the reciprocal plane $(0kl)$ at $300$ $K$ under ambient pressure in the $P2_1/m$ used by Li et al \cite{Li}($a = 5.452~\mathrm{\AA}$, $b = 5.359~\mathrm{\AA}$, $c = 10.497~\mathrm{\AA}$, $\beta=105.053^\circ$). The red arrows evidence the presence of reflections at l/2 corresponding to a doubling of the c parameter.}
\label{Fig_1}
\end{figure}

\begin{figure}[htbp]
\centering
\includegraphics[width=0.95\linewidth]{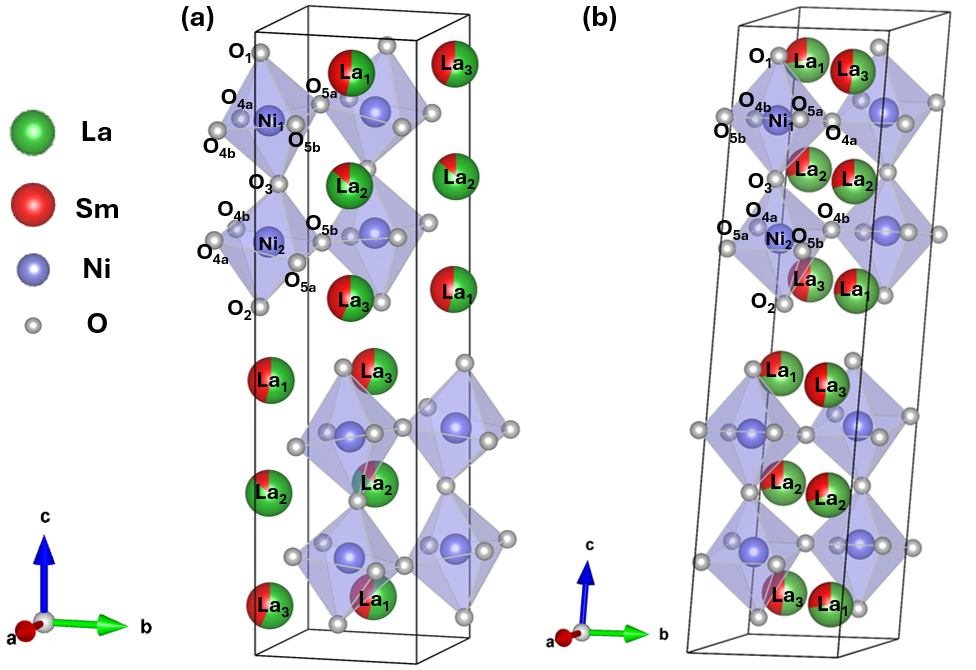}
\caption{\footnotesize(Color online) Crystal structures of \(\mathrm{La_2SmNi_2O_7}\) in the
(a) pseudo-orthorhombic \(P2_1/n\) cell ($\alpha$ = 89.9°) and
(b) monoclinic \(P2_1/c\) phase ($\beta$ $\approx$ 105°).
The blue octahedra represent the \(\mathrm{NiO_6}\) units,
while the different La/Sm mixed occupancies are illustrated
by green/red portion of spheres at different crystallographic sites.
The black lines denote the corresponding crystallographic unit cells. The atoms labels are indicated}

\label{Fig_2}
\end{figure}


\sisetup{
    separate-uncertainty = false,
    table-number-alignment = center
}


\begin{figure*}[htbp]
\centering
\includegraphics[width=0.6\linewidth]{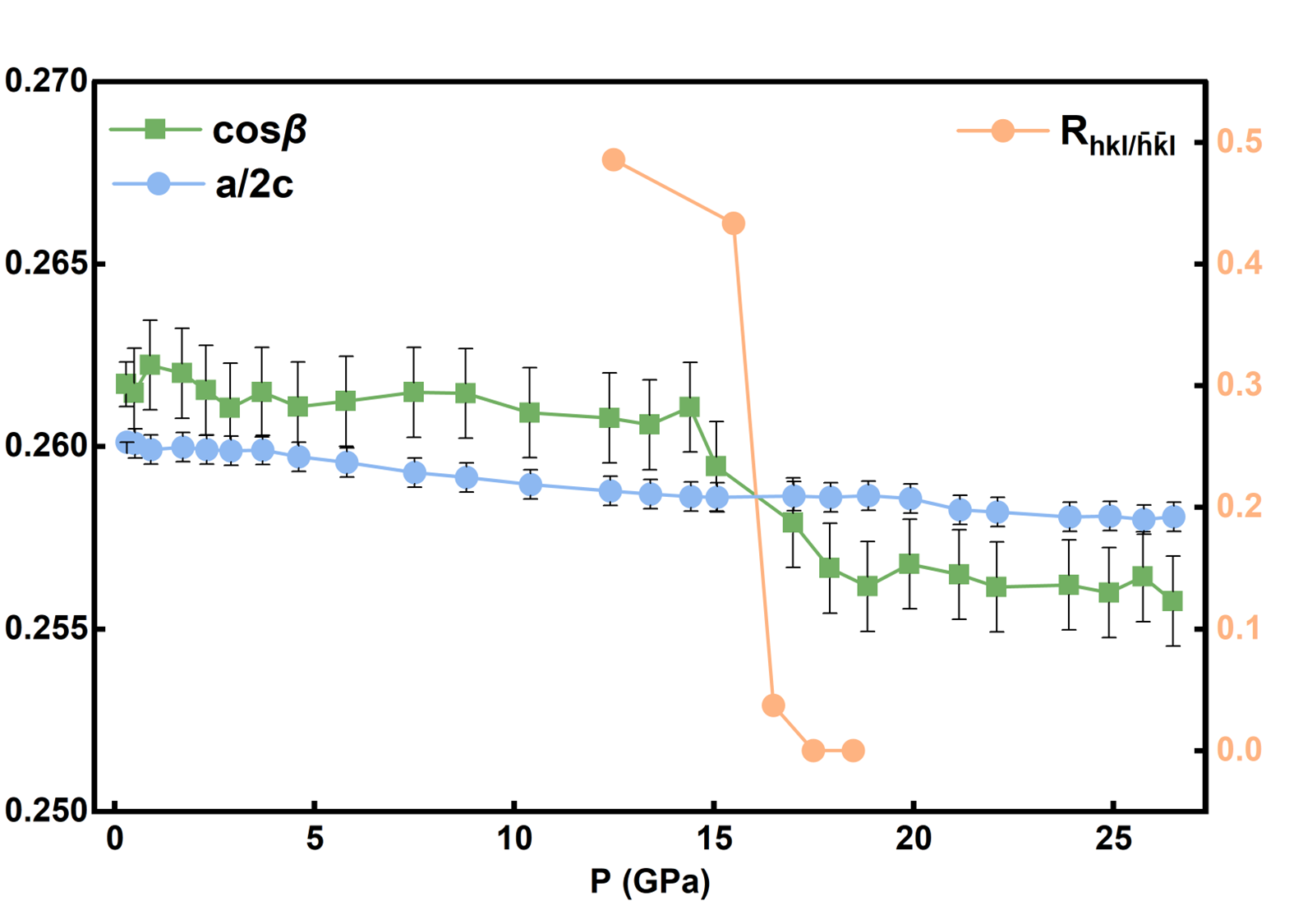}
\caption{\footnotesize(Color online) Pressure evolution of various quantities i) the $\beta$ angle of the P2$_1$/c doubled cell (green squares), ii) the $a/2c$ ratio (blue circles) and iii) the ratio R$_{hkl/\Bar{h}\Bar{k}l}$ between the number of pairs with $I_{hkl}-I_{\Bar{h}\Bar{k}l}$ greater than $3*\sigma$, divided by the total number of pairs related by the 2 fold rotation along $c$ (orange circles and scale at the right). Due to the bad quality of the data above $20$ $GPa$, the curve is interrupted.}
\label{Fig_3f}
\end{figure*}

\begin{figure}[htbp]
\centering
\includegraphics[width=0.95\linewidth]{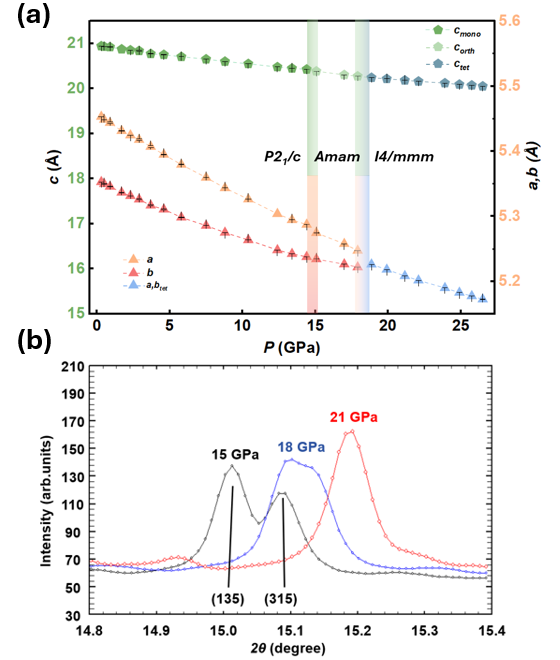}
\caption{\footnotesize(Color online) a) Pressure evolution at room temperature of the unit cell parameters. Note that the a and b unit cell parameters are inverted in the $P2_1/c$ (b unique) and $Amam$ setting but for a sake of simplicity we kept the $P2_1/c$ setting for this curve until $18$ $GPa$. Note that while the $Amam$ space group is consistent with the present data, it cannot be established unambiguously — only the point group symmetry can be determined with certainty. b) Enlarged view of the x-ray diffractogram presented in Fig. S5, representing the merge of the (135) and (315) reflections (in the $Amam$ setting) as a function of pressure}
\label{Fig_3}
\end{figure}

From our powder x-ray diffraction as a function of pressure at room temperature we were able to extract by Le Bail fit the unit cell parameters. We used the $P2_1/c$ description of the structure (i.e. $c$-doubled cell with $\beta$$\approx$105°). The pressure evolution of the unit cell parameters is shown in Fig. \ref{Fig_3}a revealing a smooth compression. Surprisingly the relative compression is isotropic even though the structure consists of quasi-two-dimensional layered system. From these measurements, we deduced the equation of state of \LSNO{} and obtain a bulk modulus of B$_0$ = 147.3 GPa (see Fig S6 of SI) very similar to the one of \LNO{} (B$_0$ = 143.6 GPa) \cite{Xu}. 

Furthermore, Fig. \ref{Fig_3f} shows the evolution of the cosinus of $\beta$ compared with $a/2c$. In a purely orthorhombic cell they should be equal. From Fig. \ref{Fig_3f}, an abrupt variation of $cos(\beta)$ is observed at $15$ $GPa$. This pressure marks the onset of a structural transition either isosymmetric or not. To assess if the structure just above $15$ $GPa$ is of higher symmetry (orthorhombic or tetragonal), we compared the intensity I, of the pairs of reflections (hkl)/($h\Bar{k}\Bar{l}$) and (hkl)/($\Bar{h}\Bar{k}l$) which are related by twofold rotation axes along $a$ and $c$ respectively, symmetry elements not present in a monoclinic setting. The data come from our single crystal measurement at $10$ $K$. In particular, we studied as a function of pressure, the ratio between the number of pairs with $I_{hkl}-I_{h\Bar{k}\Bar{l}}$ larger than $3*\sigma$ over the total number of pairs. This ratio is of about 50\% below $15$ $GPa$ and decreases to 0\% above, confirming the onset of at least an orthorhombic symmetry at this pressure. For more details see SI. The exact space group of this phase is difficult to establish from powder measurements but the x-ray pattern is consistent with the $Amam$ of the parent \LNO{} system.

At $15$ $GPa$,  Fig. \ref{Fig_3}b evidences that the (135) and (315) Bragg reflections (in the orthorhombic setting) are not equivalent confirming that the symmetry is genuinely orthorhombic rather than tetragonal at this pressure. Above $15$ $GPa$, one observes a progressive merging of these peaks  characteristic of the onset of a tetragonal symmetry. The system becomes quasi-tetragonal above $18$ $GPa$, although the transition is only fully completed at $21$ $GPa$ as the reflection's width perfectly corresponds to the experimental resolution. This result evidences that the intermediate orthorhombic phase between the monoclinic $P2_1/c$ and tetragonal $I4/mmm$, survives in the pressure range $15$ $GPa$ to $21$ $GPa$ at room temperature. Notice that this phase is absent from the recent phase diagram proposed from DFT calculations \cite{Zhao} and has not been observed in the trilayer compound either \cite{ramakrishnan2026}. As for the critical pressure of the tetragonal transition (onset at $18$ $GPa$ and completed at $21$ $GPa$), it is broadly consistent with the one reported at $18$ $GPa$ in the study of Li et al \cite{Li}. Importantly, the critical pressures in \LSNO{} are higher than the one, $12$ $GPa$, of the undoped system \cite{Puphal}. These  counterintuitive results confirm previous works whereby chemical pressure shifts the orthorhombic-to-tetragonal transition to higher pressures compared to the pristine compound \cite{Wang_pressure, osada2026}.

Above $20$ $GPa$, the sample was partially damaged, likely due to bridging between anvils. We were able to confirm both the $P2_1/n$ space group and the doubling of the $c$ axis at low pressure and $10$ $K$. However, the number of indexed reflections were insufficient to refine the structure in the doubled cell. This difficulty arises from the strong incoherent scattering (background) associated with the diamond anvil cell, which masks the weak superstructure reflections related to the c-axis doubling. 

In addition, due to the small opening of the DAC, less than 75 \% of the reflections below 0.6 \text{\AA} were collected, resulting in low completeness. Similar limitations were reported recently by Xu et al \cite{Xu}, although those
measurements were not conducted at low temperature. The structure was therefore refined only in the average $P2_1/m$ structure determined by Li et al \cite{Li}. The results for two particular pressures ($3.5$ and $18.5$ $GPa$) are given in tables S3, S4 and S5 of SI.

\begin{figure}[htbp]
\centering
\includegraphics[width=0.9\linewidth]{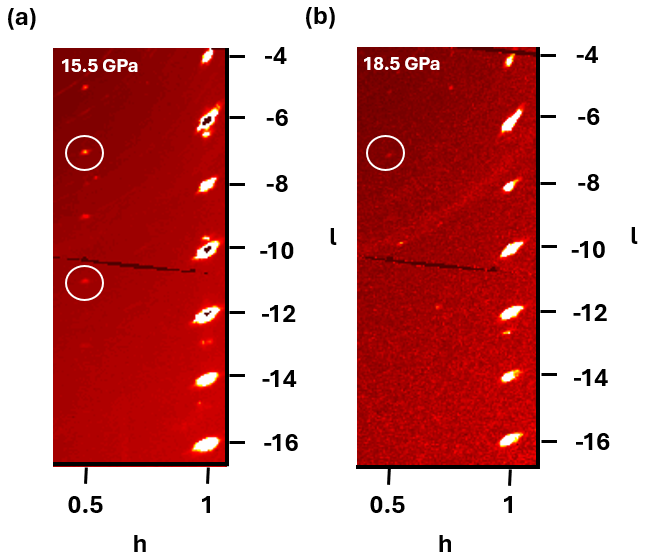}
\caption{\footnotesize(Color online) Reconstruction of one (hhl) plane in the $I4/mmm$ tetragonal setting at a) $15.5$ $GPa$ et b) $18.5$ $GPa$. Circles indicate ($\frac{1}{2}$ $\frac{1}{2}$ l) reflections characteristic of the orthorhombic lattice.}
\label{Fig_4}
\end{figure}

Fig. \ref{Fig_4} illustrates the evolution of the reciprocal space at high pressure. The monoclinic-to-orthorhombic transition cannot be well identified from this single-crystal measurement. It is however possible to check the extinction rules of the $Amam$ space group. Indications of the orthorhombic-to-tetragonal transition are more visible. They are evidenced in the (hhl) reconstruction of the $I4/mmm$ reciprocal space of Fig. \ref{Fig_4}. They are characterized by the progressive disappearance upon pressure of ($\frac{1}{2}$ $\frac{1}{2}$ l) reflections (indexed in the tetragonal lattice setting) which are characteristic of the $\sqrt{2}a \times \sqrt{2}b \times c$ orthorhombic lattice. Above $18.5$ $GPa$, these reflections—even present— are weak so the structure was refined already in the tetragonal $I4/mmm$ space group. The refinement results are provided in the supplementary information. 


From these refinements, we were able to track the pressure evolution of the apical Ni–Ni distances as well as the Ni-O-Ni angles along the $c$ direction (see Fig. S7 of SI). Within the accuracy of the refinements, the Ni-Ni distance along $c$ remains independent of the pressure. This can be explained by the action of two opposite effects : the pressurization which is expected to reduce the unit cell parameters and thus the atomic distances and decrease the octahedra tilt relatively to the c-axis as the Ni-O-Ni angle tends to be linear. As for the Ni–O–Ni angle along $c$, it increases monotonously from 162$\degree$ at ambient pressure to 180$\degree$ at $21$ $GPa$ in the tetragonal phase. The details of the refinements for two pressures are given in Tables S3 to S4 of the SI.    

Our findings shed light on three important features via the determination of structural properties at carefully selected points of the (P,T) phase diagram. 

First, we report a departure from the published ambient-pressure structure, previously determined using laboratory X-ray sources or powder diffraction. Indeed, our synchrotron single-crystal measurements reveal a doubling of the $c$ parameter, arising from an antiferrodistortive-like pattern of the NiO$_6$ octahedra driven by displacements of the O4a, O4b and O5a, O5b oxygen atoms constituting the conducting Ni–O bilayers. As illustrated in Fig. S8, these alternating displacements from NiO bilayer to NiO bilayer are predominantly along $c$ for O5a and O5b ($\Delta$z=0.22 $\AA$) and within the (a,b) plane for O4a and O4b ($\Delta$xy=0.06 $\AA$). A similar antiferrodistortive phase is known in perovskite systems \cite{SrTIO} and has also been recently predicted for infinite-layer nickelate superconductors \cite{InfiniteLayer}. This structural feature has a significant impact on electronic properties not only because it directly modulates the $dx^2$-$y^2$ transfer integrals but also for its effect on the nematic phase which has been recently evidenced \cite{NematicLaSm}. Indeed, in the P2$_1$/c monoclinic phase, not only the x,y directions within the NiO layer are anisotropic due to the breaking of the c4 rotation axis lost at the tetragonal to orthorhombic (or monoclinic) transition, but also their diagonals are no more equivalent due to the antiferrodistortive-like deformation. This is expected to modify the possible directors of the  electronic nematic phase \cite{Nematic} . 


Second, we demonstrate the absence of any sizeable displacive CDW in \LSNO{}, consistent with findings in pristine \LNO{} \cite{Khasanov2025}. Given the sensitivity of our measurements, any CDW-related atomic displacement would be smaller than a few thousandths of an \text{\AA}\, weaker than that observed in canonical CDW systems where the modulation is rather $5*10^{-2}$ \text{\AA}\ \cite{Bronzebleu,Tsunoda}. Such a vanishingly small displacive amplitude, points either to weak electron–phonon coupling or to poor Fermi surface nesting. Alternatively, it suggests a purely electronic CDW, analogous to the $2k_F$ charge instability observed in the organic superconductor TMTSF$_2$PF$_6$ \cite{Pouget}, or to hole stripes in cuprates \cite{Tranquada}. Another option could be due to the disorder related to the La/Sm occupancy of the rare earth sites which could impact the CDW ordering even if it is located out of the Ni layer. On the contrary in n = 3 Ruddlesden–Popper (RP) nickelates, a sizeable displacive CDW coupled to the SDW via exchange-striction has been reported \cite{L4NO,ramakrishnan2026}. 
This contrast is particularly striking given that the Fermi surface of n = 2 RP nickelates is expected to be flatter than that of n = 3 compounds — a geometry generally more favorable to nesting-driven density-wave instabilities. This difference may instead be attributed to the smaller SDW amplitude reported in n = 2 compounds \cite{plokhikh2025}, which could reflect weaker Coulomb interactions or electron–phonon coupling, both key parameters controlling SDW instabilities with or without CDW harmonics \cite{Overhauser,Young_1974}. It is indeed known for both n = 2 and n = 3 systems that the electron-phonon coupling is weak \cite{michon2026,suthar2025}. An additional factor can be a reduced bulk modulus known to govern the strain-wave mechanism as established in canonical SDW metals such as chromium \cite{Tsunoda, Fawcett}. Since the bulk moduli (SI and \cite{ramakrishnan2026}) and electron–phonon coupling \cite{KUMAR,JZan2025} of n=2 and 3 RP nickelates are comparable, one can infer that it is electron–electron interactions which are intrinsically weaker in the n=2 family.

Third, through systematic powder x-ray measurements under varying pressure at room temperature, we identified a sequence of structural phase transitions: from monoclinic to orthorhombic at $15$ $GPa$, and subsequently to tetragonal at $21$ $GPa$ (onset already at $18$ $GPa$). Combining our structural results with the transport measurements of Li et al. \cite{Li}, which were performed on samples of identical synthesis and are therefore expected to hold for the present crystals, we find that the onset of the superconducting dome ($\approx$ $15$ $GPa$) exactly coincides with the disappearance of the monoclinic phase in absence of strictly linear Ni-O-Ni angle along $c$. This provide strong evidence of the incompatibility of monoclinic symmetry with the SC phase. However T$_c$ reaches its maximum within the tetragonal $I4/mmm$ phase which confirms the importance of a linear angle to optimize the SC properties as recently evidenced theoretically \cite{Lecherman}. Indeed this work demonstrated that the SDW phase is destabilized by the tetragonal symmetry which thus favors SC. More generally, under pressure, the hybridization between the O p and Ni d orbitals is significantly enhanced due to the increase in the Ni–O–Ni bond angle. As a result, the relative occupancies of the $dx^2$-$y^2$ and $dz^2$ orbitals change substantially. This orbital redistribution is a key ingredient that any minimal model of Cooper pairing must capture. To conclude, we provide here structural parameters across the superconducting transition and its precursor regime. These structural information can be used for further DFT calculations and modeling of SC transition. 

\begin{acknowledgments}
The authors thank SOLEIL for synchrotron beam time on CRISTAL beamline (Proposal N° 20251520 and 20251514). This work was financially supported by the ANR SUNSHINE No.ANR-25-CE30-7264 and SUPERNICKEL No.ANR-21-CE30-0041-04 as well as European Research Council (ERC) under the European Union’s Horizon 2020530 research and innovation program (Grant Agreement No. 865826) and the Paris Ile-de-France Region in the framework of DIM MaTerRE (project DAC-VX). We also thank University of Bordeaux for the “Invited Professor” position awarded by Sourav Marik through the Visiting Scholar Program
\end{acknowledgments}

\bibliographystyle{apsrev4-2}
\bibliography{Biblio}

\end{document}


\title{Supplementary Information for Pressure-Driven Structural Transitions without a Displacive Charge-Density Wave in \LSNO}

\author{J. Huang}
\affiliation{Universit\'e Paris-Saclay, CNRS, Laboratoire de Physique des Solides, 91405, Orsay, France.}
\affiliation{Synchrotron SOLEIL, L\'\ Orme des Merisiers, Saint Aubin BP 48, 91192, Gif-sur-Yvette, France}

\author{Sitaram Ramakrishnan}
\affiliation{CNRS, Universit\'e Grenoble Alpes, Institut N\'eel, F-38000 Grenoble, France}

\author{P. Rodi\`ere}
\affiliation{CNRS, Universit\'e Grenoble Alpes, Institut N\'eel, F-38000 Grenoble, France}

\author{P. Toulemonde}
\affiliation{CNRS, Universit\'e Grenoble Alpes, Institut N\'eel, F-38000 Grenoble, France}

\author{Z. Rahmany}
\affiliation{Universit\'e Paris-Saclay, CNRS, Laboratoire de Physique des Solides, 91405, Orsay, France.}

\author{V. Bal\'edent}
\affiliation{Universit\'e Paris-Saclay, CNRS, Laboratoire de Physique des Solides, 91405, Orsay, France.}
\affiliation{Institut universitaire de France (IUF)}

\author{B. Vignolle}
\affiliation{Institut de Chimie de la mati\`ere condens\'ee de Bordeaux, Universit\'e Bordeaux, France}

\author{S. Marik}
\affiliation{Institut de Chimie de la mati\`ere condens\'ee de Bordeaux, Universit\'e Bordeaux, France}

\author{P. Fertey}
\affiliation{Synchrotron SOLEIL, L\'\ Orme des Merisiers, Saint Aubin BP 48, 91192, Gif-sur-Yvette, France}

\author{P. Foury-Leylekian}
\affiliation{Universit\'e Paris-Saclay, CNRS, Laboratoire de Physique des Solides, 91405, Orsay, France.}
\email [Corresponding author: ] {pascale.foury@universite-paris-saclay.fr}

\maketitle

\onecolumngrid

\noindent\textbf{1. Reciprocal-space reconstruction and structural refinement of $\mathrm{La_2SmNi_2O_7}$ at ambient pressure}

To clarify the difference between our results and the previous work of Li et al \cite{Li}, reciprocal-space reconstructions were performed at ambient pressure. Fig. S1 shows the reconstruction at $15$ $K$. We observe weak reflections at half-integer $l$ positions in the (0kl) reconstruction, indicating a doubling of the $c$ axis relative to the previously reported primitive monoclinic cell \cite{Li}. 

The reconstruction of the (h0l) reciprocal plane at $300$ $K$ shown in Fig. S2 evidences the extinction rules $l=2n$ characteristic of a $c$ glide mirror. This leads to the conclusion that the space group is $P2_1/c$.

Fig. S3 shows the absence of additional reflections at low temperature compared to $300$ $K$. No CDW satellite reflection is detected. 

Fig. S4 summarizes the relationship between the different settings employed throughout this work. Blue and black were used to differentiate the $P2_1/m$ double-c monoclinic cell observed in our work from the previously reported $P\,1\,2_1/m\,1$ simple-c monoclinic cell of Li et al \cite{Li}. In addition, the quasi-orthorhombic $P2_1/n$ cell is shown in red, it is nearly equivalent to an orthorhombic one due to the particular value of $\beta$.

\begin{figure*}[htbp]
\centering
\includegraphics[width=0.6\linewidth]{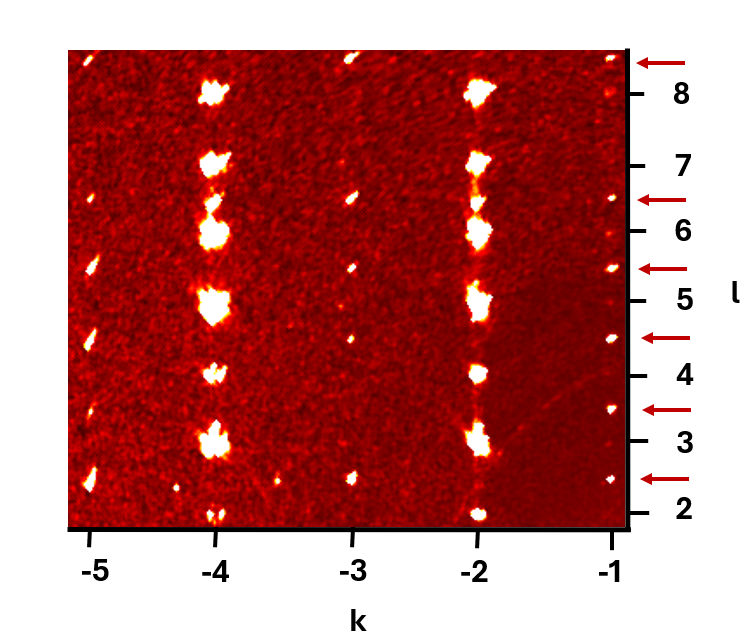}
\caption{\footnotesize(Color online) Reconstruction of the (0kl) reciprocal plane in the setting of reference \cite{Li} at $15$ $K$. }
\label{Fig_S1}
\end{figure*}

\begin{figure*}[htbp]
\centering
\includegraphics[width=0.6\linewidth]{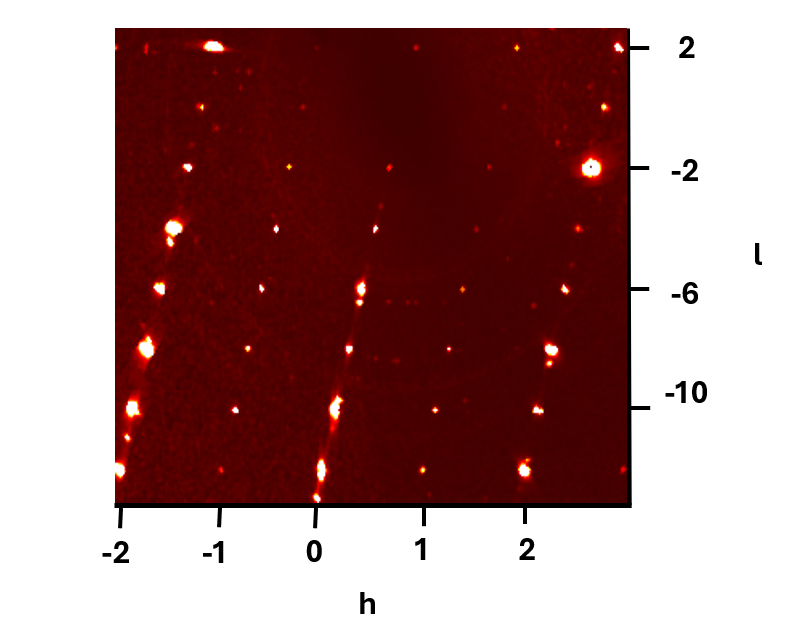}
\caption{\footnotesize(Color online) Reconstruction of the (h0l) reciprocal plane in the doubled-c monoclinic setting at $300$ $K$. }
\label{Fig_S1}
\end{figure*}

\begin{figure*}[htbp]
\centering
\includegraphics[width=1\linewidth]{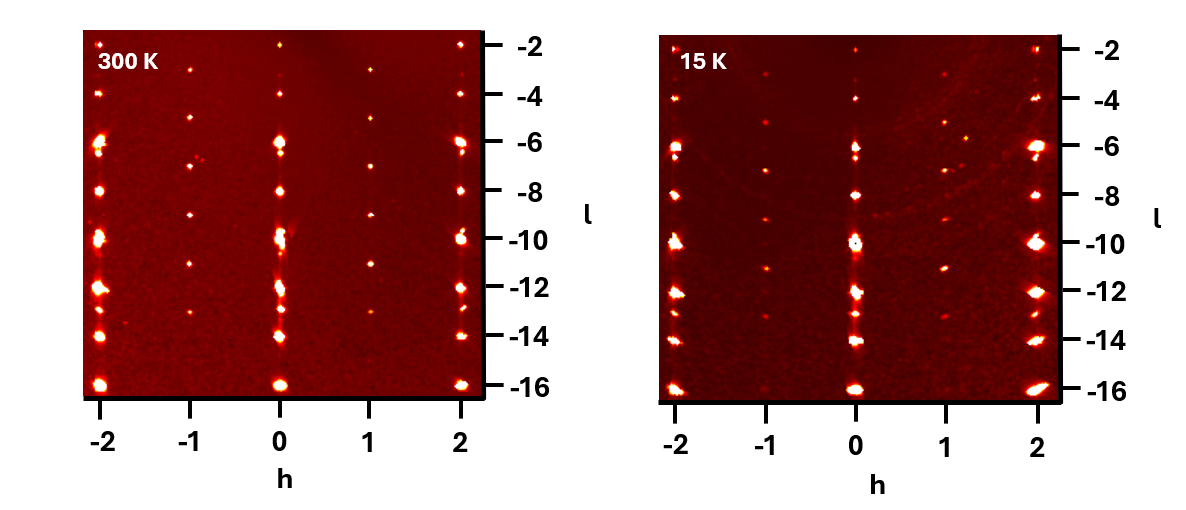}
\caption{\footnotesize(Color online) Reconstruction of the $(h0l)$ plane of the $P2_1/n$ pseudo-orthorhombic reciprocal space at ambient pressure for $300$ $K$ (left) and $15$ $K$ (right). No additional reflection appears at low temperature}
\label{Fig_S2}
\end{figure*}

\begin{figure*}[htbp]
\centering
\includegraphics[width=0.8\linewidth]{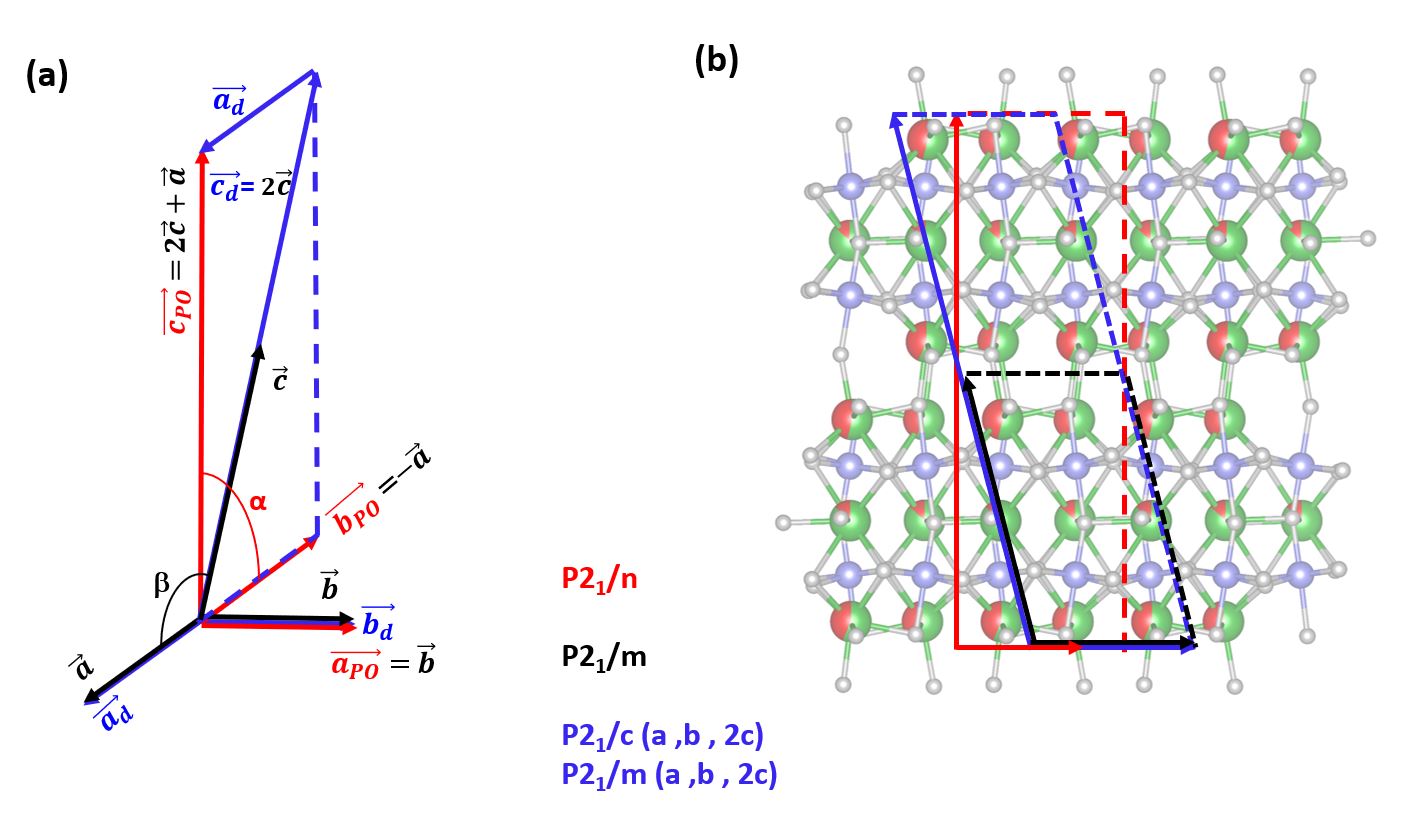}
\caption{\footnotesize(Color online) Schematic representation of the different unit-cell descriptions used in this work in perspective (a) and in a 2D projection (b): 
the initial monoclinic $P2_1/m$ structure reported by Li \textit{et al.} with a, b, c (black); 
the doubled-$c$ monoclinic $P\,1\,2_1/m\,1$ and $P\,1\,2_1/c\,1$ cells with a$_{d}$,b$_{d}$,c$_{d}$ (blue); 
and the equivalent $P2_1/n\,1\,1$ pseudo-orthorhombic (PO) description with a$_{PO}$,b$_{PO}$,c$_{PO}$ (red). 
The $P2_1/n$ pseudo-orthorhombic structure becomes orthorhombic for $\cos(\beta)=a/2c$.}
\label{Fig_2}
\end{figure*}

 \begin{table*}[t]
\centering
\caption{Fractional atomic coordinates and thermal displacement parameters (\AA$^2$) in the $P2_1/n$ space group at $15$ $K$ and ambient pressure with a = 5.3639(5) \text{\AA}\ b = 5.4365(4) \text{\AA}\ c = 20.2154(17) \text{\AA}\ $\alpha$ = 89.926$\degree$ $R_{obs}$ = 7.61 \% . The anisotropic ADP factors of oxygen and $U_{23}$ for any atom are not refined.}
\label{tab:structure}

\begin{tabular*}{\textwidth}{
@{\extracolsep{\fill}}c cccc cccccc}
\hline

& \multicolumn{4}{c}{Coordinates} & \multicolumn{6}{c}{Displacement parameters} \\
\cline{2-11}

Atom & $occ$ & $x$ & $y$ & $z$ & $U_{11}$/$U_{iso}[O]$ & $U_{22}$ & $U_{33}$ &
$U_{12}$ & $U_{13}$ & $U_{23}$ \\
\hline

La1 & 0.54(3) &	-0.2516(3)& 0.0125(2)& 0.43068(6)&	0.0080(5)& 0.0079(5)& 0.0012(4)& -0.0034(2)& -0.00137(19)& -0.000008

 \\
Sm1 & 0.46(3) &	-0.2516(3)& 0.0125(2)& 0.43068(6)&	0.0080(5)& 0.0079(5)& 0.0012(4)& -0.0034(2)& -0.00137(19)& -0.000008
 \\
La2 & 0.89(3)&	-0.2519(3)& 0.00287(12)& 0.25001(7)&	 0.0069(5)& 0.0074(5)& 0.0023(4)& 0.0008(3)& -0.0008(2)& -0.000008
 \\
Sm2 & 0.11(3)&	-0.2519(3)& 0.00287(12)& 0.25001(7)&	 0.0069(5)& 0.0074(5)& 0.0023(4)& 0.0008(3)& -0.0008(2)& -0.000008
 \\
La3 & 0.58(3)&	-0.2488(3)& 0.0126(2)& 0.06936(6)&	0.0080(5)& 0.0079(5)& 0.0012(4)& -0.0034(2)& -0.00137(19)& -0.000008

 \\
Sm3 & 0.42(3)&	-0.2488(3)& 0.0126(2)& 0.06936(6)&	0.0080(5)& 0.0079(5)& 0.0012(4)& -0.0034(2)& -0.00137(19)& -0.000008
\\
Ni1 & 1&	-0.2519(6)& 0.5124(4)& 0.34748(13)	&0.0034(3)&	-&	-& - & - & -
 \\
Ni2 & 1&	-0.2519(6)& 0.4951(4)& 0.15274(14)&	0.0034(3)&	-&	-& - & - & - \\
O1  & 1&	-0.271(3))& 0.454(4) &0.4518(7)&	0.0062(16)&	-&	-& - & - & -
 \\
O2  & 1&	-0.725(3)& 0.052(3) &0.4588(7)&	0.0062(16)&	-&	-& - & - & - \\
O3  & 1&	-0.233(3)& 0.556(2)& 0.2488(9)&	0.0065(18)&	-&	-& - & - & - \\
O4a  & 1&	-1.004(4)& 0.265(3) &0.3377(4)&	0.0007(18)&	-&	-& - & - & -
 \\
O4b  & 1&	-0.973(3)& -0.234(4)& 0.8386(6)&	0.012(3)&	-&	-& - & - & -
 \\
 
O5a  & 1&	0.013(4)& -0.270(3)& 0.3662(6)&	0.0061(15)&	-&	-& - & - & -\\
O5b  & 1&	-0.021(3) &0.249(3) &0.8545(5)&	0.0061(15)&	-&	-& - & - & -
 \\

\hline
\end{tabular*}
\end{table*}

 \begin{table*}[t]
\centering
\caption{Fractional atomic coordinates and thermal displacement parameters (\AA$^2$) in the $P2_1/n$ space group at 300 K and ambient pressure with a = 5.3551(3) \text{\AA}\ b = 5.4513(3) \text{\AA}\ c = 20.2557(5) \text{\AA}\ $\alpha$ = 89.932$\degree$ $R_{obs}$ = 6.10 \% . The quality of the refinement is artificially better than at $15$ $K$ due to the smaller number of weak reflections recorded at $300$ $K$. The anisotropic ADP factors of oxygen and $U_{23}$ for any atom are not refined.}
\label{tab:structure}

\begin{tabular*}{\textwidth}{
@{\extracolsep{\fill}}c cccc cccccc}
\hline

& \multicolumn{4}{c}{Coordinates} & \multicolumn{6}{c}{Displacement parameters} \\
\cline{2-11}

Atom & $occ$ & $x$ & $y$ & $z$ & $U_{11}$/$U_{iso}[O]$ & $U_{22}$ & $U_{33}$ &
$U_{12}$ & $U_{13}$ & $U_{23}$ \\
\hline

La1 & 0.54(3) &	-0.2518(4)& 0.0150(3)& 0.43068(3)&	0.0107(4)& 0.0093(4)& 0.0030(4)& -0.0032(5)& 0.0036(4)& 0.000741
 \\
Sm1 & 0.46(3) &	-0.2518(4)& 0.0150(3)& 0.43068(3)&	0.0107(4)& 0.0093(4)& 0.0030(4)& -0.0032(5)& 0.0036(4)& 0.000741
 \\
La2 & 0.89(3)&	-0.2460(6)& 0.00264(13)& 0.25037(10)&	 0.0103(4)& 0.0112(4)& 0.0044(4)& -0.0027(6)& 0.0012(5)& 0.000741
 \\
Sm2 & 0.11(3)&	-0.2460(6)& 0.00264(13)& 0.25037(10)&	 0.0103(4)& 0.0112(4)& 0.0044(4)& -0.0027(6)& 0.0012(5)& 0.000741
 \\
La3 & 0.58(3)&	-0.2520(4)& 0.0090(3)& 0.06932(3)&	0.0107(4)& 0.0093(4)& 0.0030(4)& -0.0032(5)& 0.0036(4)& 0.000741
 \\
Sm3 & 0.42(3)&	-0.2520(4)& 0.0090(3)& 0.06932(3)&	0.0107(4)& 0.0093(4)& 0.0030(4)& -0.0032(5)& 0.0036(4)& 0.000741 \\

Ni1 & 1&	-0.2392(10)& 0.50369(19)& 0.34730(7)&	0.0045(3)&	-&	-& - & - & -

 \\
Ni2 & 1&	-0.2560(10)& 0.50369(19)& 0.15270(7)&	0.0045(3)&	-&	-& - & - & -

 \\
O1  & 1&	-0.246(5)& 0.466(4) &0.4524(14)&	0.020(5)&	-&	-& - & - & -
 \\
O2  & 1&	-0.722(3)& 0.060(3)& 0.4574(10)
&	0.001(3)&	-&	-& - & - & - 
\\
O3  & 1&	-0.236(4)&0.555(2) &0.2497(17) &	0.011(2)&	-&	-& - & - & - \\
O4a  & 1&	-1.006(5)& 0.268(4)&0.3439(7)&	0.005(2)&	-&	-& - & - & -
 \\
O4b  & 1&	-1.021(4)& -0.261(4)& 0.8340(7)&	0.005(2)&	-&	-& - & - & -
 \\
O5a  & 1&	-0.001(5)&-0.226(4)& 0.3664(7)&	0.005(2)&	-&	-& - & - & -\\
O5b  & 1&	-0.014(6)&0.241(5) &0.8526(7)&	 0.005(2)&	-&	-& - & - & -\\

\hline
\end{tabular*}
\end{table*}

\clearpage
\noindent\textbf{2.Pressure evolution of the $\mathrm{La_2SmNi_2O_7}$ crystal structure from powder X-ray diffraction}

\begin{figure*}[htbp]
\centering
\includegraphics[width=0.6\linewidth]{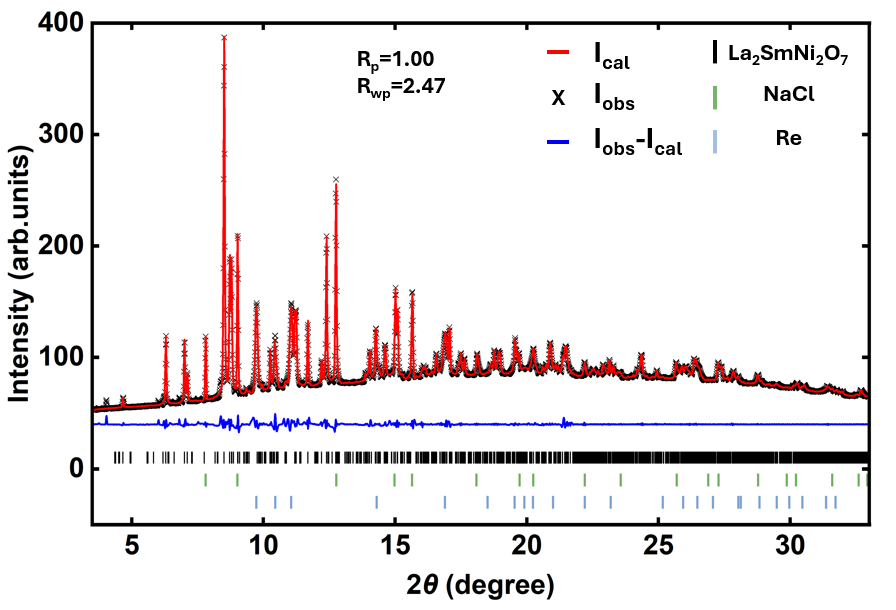}
\caption{\footnotesize(Color online)  Observed (black fork), calculated (red line), and difference (dark blue line) profiles obtained from the La$_2$SmNi$_2$O$_7$ at 15 GPa of the Rietveld refinement.The tick marks represent the Bragg peak positions of La$_2$SmNi$_2$O$_7$ (black), NaCl (green), and Re (blue), respectively..
}
\label{Fig_3}
\end{figure*}

Powder diffraction data for \LSNO{} (Fig. S5) allows us to track the evolution of unit cell parameters as a function of pressure. Fig. S6 shows the evolution of $\cos\beta$ and $a/2c$ as a function of pressure.
The crossing between both curves at $15$ $GPa$ suggests a transition from the monoclinic $P2_1/c$ structure to an orthorhombic one at this pressure. 

We compared the intensity of the pairs of reflections (hkl)/($h\Bar{k}\Bar{l}$) and (hkl)/($\Bar{h}\Bar{k}l$) which are related by twofold rotation axes along $a$ and $c$ characteristic of the orthorhombic symmetry. At $12$ $GPa$ upon the 70 pairs related by the 2 axis along $a$ ($c$), 35 (4) exhibit a significant difference of intensity with an average ratio $I_{hkl}/I_{h\Bar{k}\Bar{l}}(I_{\Bar{h}\Bar{k}l})$ of approximately $5*\sigma$. At $18$ $GPa$, only 2 pairs display a measurable intensity difference but which falls within the error bars. We also present in Fig. S6, the pressure evolution of the ratio between the number of pairs with $I_{hkl}/I_{h\Bar{k}\Bar{l}}$ stronger than $3*\sigma$ over the total number of pairs related by the 2 fold axis along $c$. These results evidence a monoclinic-orthorhombic transition at $15$ $GPa$.


Fig. S6 represents the fit of the powder data under pressure using the third-order Birch--Murnaghan equation of state. 
\begin{equation}
P=\frac{3}{2}K_0
\left[
\left(\frac{V_0}{V}\right)^{7/3}
-
\left(\frac{V_0}{V}\right)^{5/3}
\right]
\left\{
1+
\frac{3}{4}(K_0'-4)
\left[
\left(\frac{V_0}{V}\right)^{2/3}-1
\right]
\right\}
\end{equation}
\begin{figure*}[htbp]
\centering
\includegraphics[width=0.6\linewidth]{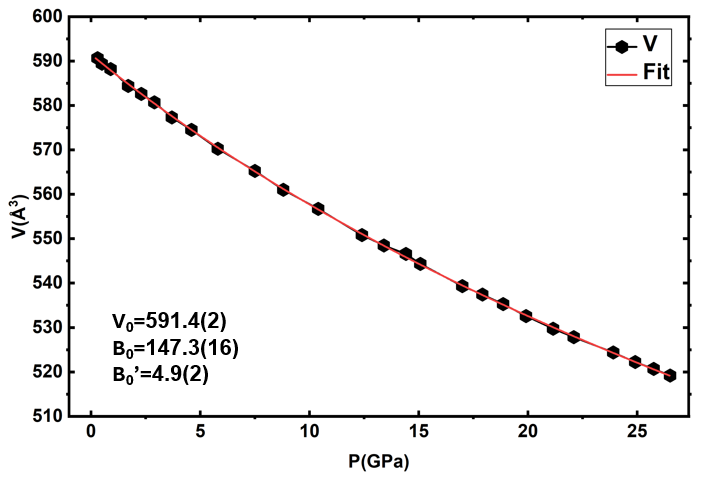}
\caption{\footnotesize(Color online) Pressure dependence of the unit-cell volume. The solid red line represents the fit using the third-order Birch--Murnaghan equation of state, yielding 
$V_0 = 591.4(2)$ \AA$^3$ and $B_0 = 147.3(16)$ GPa with the pressure derivative $B_0^{\prime}=4.9$(2))}
\label{Fig_S7}
\end{figure*}

\clearpage
\noindent\textbf{3.Pressure evolution of structural refinements and unit-cell parameters in \LSNO from single-crystal X-ray diffraction}

\begin{figure*}[htbp]
\centering
\includegraphics[width=0.6\linewidth]{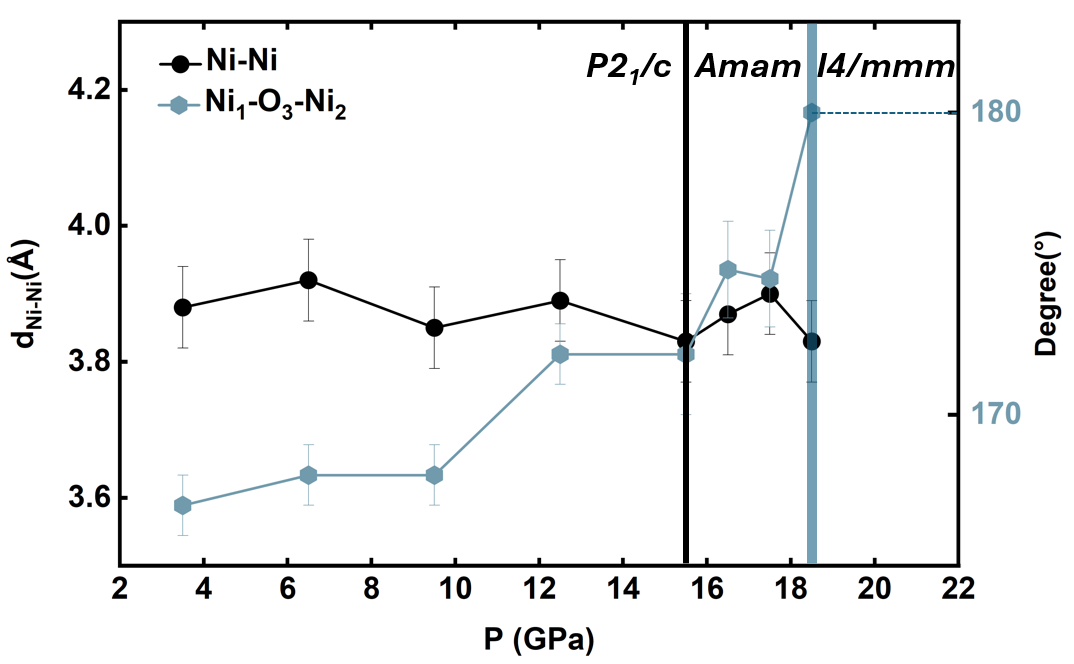}
\caption{\footnotesize(Color online) Pressure evolution of the Ni-Ni apical distance and Ni-O-Ni angle along $c$. Note that the $Amam$ to $I4/mmm$ transition is totally achieved only at $21$ $GPa$}
\label{Fig_S6}
\end{figure*}

\begin{figure*}[htbp]
\centering
\includegraphics[width=0.6\linewidth]{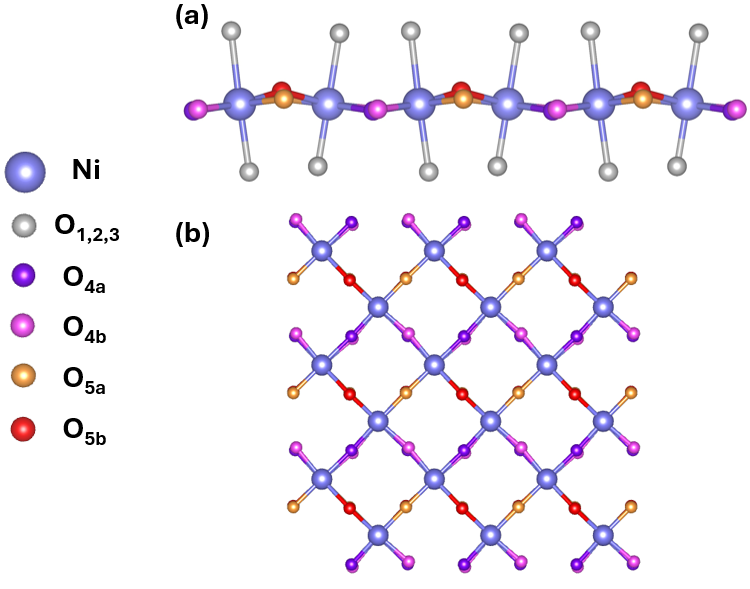}
\caption{\footnotesize(Color online) (a) Projection along $a$ of one NiO$_6$ plane at $15$ $K$ in the $P2_1/n$ structure (b) Projection along $c$ of one NiO$_6$ plane at $15$ $K$ in the $P2_1/n$ structure .}
\label{Fig_S6}
\end{figure*}

Fig. S7 represents the evolution and various structural parameters as a function of pressure. 
Tables S3 and S4 summarize the refinement details for the low-pressure simple-c monoclinic $P2_1/m$ phase and the high-pressure tetragonal $I4/mmm$ phase, respectively.

The difference between the $P2_1/m$ (a,b,c) and $P2_1/c$ (a,b,2c) structures is subtle, since most atoms lie on the mirror plane perpendicular to the unique axis. As a result, these atoms are exactly related by a c/2 translation in the doubled-c cell and therefore remain crystallographically equivalent just like in the $P2_1/m$ lattice.

Only two types of oxygen atoms in the $P2_1/m$ (a,b,c) structure, O4 and O5, are located away from the mirror plane. Through the m-symmetry operation, they generate four oxygen positions, denoted O4a, O4b, O5a, and O5b. In the $P2_1/c$ structure, these four oxygen sites become independent and each gives rise to an additional symmetry-related position through the c-glide operation. The resulting set of oxygen atoms is responsible for the doubling of the c lattice parameter.

Fig. S8 presents a schematical view of one NiO layer involving the O4a,b and O5a,b oxygen atoms. It evidences the presence of two equilibrium positions for these atoms. For example in the case of Fig. S8, the two positions are O5a and O5b : the first atom has a z coordinate smaller than the second one ($\Delta$z=0.22 $\AA$). If one considers the 2 NiO bilayers of the double-c cell, the configuration of the O5a,b doublets are identical in the first bilayer while for the second one they are opposite. This leads to an antiferrodistortive-like motif of O5a,b positions along $c$ which explains the doubling of the $c$ axis. As for the displacements of the O4a,b oxygens, they are mainly in the (a,b) plane but arranged in the same antiferrodistortive-like way from NiO bilayer to the next as for the O5ab.

\begin{table*}[h]
\centering
\caption{Fractional atomic coordinates and thermal displacement parameters (\AA$^2$) in the $P2_1/m$ space group at $10$ $K$ under $3.5$ $GPa$ with a=5.3840(18) \text{\AA}\ b=5.2852(16) \text{\AA}\ c=10.336(2) \text{\AA}\ $\beta$=104.88(2)$\degree$ $R_{obs}$=7.77\%. The uncertainties given correspond to the numerical ones extracted from the Jana software. The anisotropic ADP factors of oxygen are not refined.}
\label{tab:structure}

\begin{tabular*}{\textwidth}{
@{\extracolsep{\fill}}c cccc cccccc}
\hline

& \multicolumn{4}{c}{Coordinates} & \multicolumn{6}{c}{Displacement parameters} \\
\cline{2-11}

Atom & $occ$ & $x$ & $y$ & $z$ & $U_{11}$/$U_{iso}[O]$ & $U_{22}$ & $U_{33}$ &
$U_{12}$ & $U_{13}$ & $U_{23}$ \\
\hline

La1 & 0.68(6)&	0.5559(18)&	0.75&	0.1391(14)&	0.050(8)&	0.023(10)&	0.076(15)&0& 0.036(9)& 0

 \\
Sm1 & 0.32(6)&	0.5559(18)&	0.75&	0.1391(14)&	0.050(8)&	0.023(10)&	0.076(15)&0& 0.036(9)& 0

 \\
La2 & 0.81(5)&	0.752(3)&	-0.25&	0.5016(18)&	0.068(9)&	0.014(7)& 0.12(2)& 0& 0.063(11)& 0

 \\
Sm2 & 0.19(5)&	0.752(3)&	-0.25&	0.5016(18)&	0.068(9)&	0.014(7)& 0.12(2)& 0& 0.063(11)& 0

 \\
La3 & 0.52(6)&	0.077(2)&	0.25&	0.1380(15)&0.119(9)& 0.057(12)&  0.014(11)& 0& 0.015(8)& 0

 \\
Sm3 & 0.48(6)&	0.077(2)&	0.25&	0.1380(15)&0.119(9)& 0.057(12)&  0.014(11)& 0& 0.015(8)& 0
 \\
Ni1 & 1	&0.151(2)& 0.75& 0.3035(13)&	0.034(8)& 0.011(6)&0.031(8) & 0& 0.021(8)& 0
 \\
Ni2 & 1&	0.6600(18)& 0.25& 0.3065(10)&	0.011(6)& 0.008(5)& 0.021(7)& 0& 0.009(6)& 0
 \\
O1  & 1&	0.047(9)& 0.75&0.069(6)&	0.013(5)&	-&	-& - & - & -

 \\
O2  & 1&	0.497(10)& 0.25&0.093(6)&	0.013(5)&	-&	-& - & - & -
 \\
O3  & 1&	0.230(11)& 0.75& 0.531(9)&	0.044(12)&	-&	-& - & - & -
\\
O4  &1&	 0.394(5)&-0.030(6)&0.318(2)&	0.012(4)&	-&	-& - & - & -

 \\
O5  & 1&	0.920(6)& -0.014(7)& 0.289(2)&	0.017(5)&	-&	-& - & - & -
 \\

\hline
\end{tabular*}
\end{table*}

\begin{table*}[h]
\centering
\caption{Fractional atomic coordinates and thermal displacement parameters (\AA$^2$) in the $I4/mmm$ space group at $10$ $K$ under $18.5$ $GPa$ with a=3.7036(6) \text{\AA}\ b=3.7036(6) \text{\AA}\ c=19.537(7) \text{\AA}\  $R_{obs}$=5.69 \%. The uncertainties given correspond to the numerical ones extracted from the Jana software. The anisotropic ADP factors of oxygen are not refined.}
\label{tab:structure}

\begin{tabular*}{\textwidth}{
@{\extracolsep{\fill}}c cccc cccccc}
\hline

& \multicolumn{4}{c}{Coordinates} & \multicolumn{6}{c}{Displacement parameters} \\
\cline{2-11}

Atom & $occ$ & $x$ & $y$ & $z$ & $U_{11}$/$U_{iso}[O]$ & $U_{22}$ & $U_{33}$ &
$U_{12}$ & $U_{13}$ & $U_{23}$ \\
\hline

La1 & 0.60(4)&	0&	0&	0.32005(14)&	0.0056(8)& 0.0056(8)& 0.0055(14)& 0& 0& 0
 \\
Sm1 & 0.40(4)&	0&	0&	0.32005(14)&	0.0056(8)& 0.0056(8)& 0.0055(14)& 0& 0& 0
 \\
La2 & 0.81(5)&	0&	0&	0.5&	0.0048(11) &0.0048(11)& 0.012(2)& 0& 0& 0

 \\
Sm2 & 0.19(5)&	0&	0&	0.5&	0.0048(11) &0.0048(11)& 0.012(2)& 0& 0& 0

 \\
Ni & 1&	0&	0&	0.0974(3)&	0.001(2)& 0.001(2)& 0.011(4)& 0& 0& 0
 \\
O1  & 1&	0&	0.5&	0.9009(13)&	0.018(6)&	-&	-& - & - & -

 \\
O2  & 1&	0&	0&	0.795(3)&	0.034(13)&	-&	-& - & - & -
 \\
O3  & 1&	0&	0&	0	&0.016(11)&	-&	-& - & - & -
 \\

 \hline
\end{tabular*}
\end{table*}

\begin{table*}[t]
\caption{Crystallographic data of \LSNO for pressure at 3.5 GPa and 18.5 GPa at 10 K. For all data, the anisotropic ADPs of oxygen could not be refined successfully, hence they are constrained to isotropic.}
\centering
\begin{tabular}{ccc}
\hline\hline
 & 3.5 GPa &  18.5  \\
\hline

Crystal system & Monoclinic & tetragonal \\
Laue symmetry & $2/m$& $4/mmm$ \\
Space group & $P2_1/m$ & $I4/mmm$ \\
No. & 11 & 139 \\

$a$ (\AA) & 5.3840(18) & 3.7036(6)  \\
$b$ (\AA) & 5.2852(16 & 3.7036(6)  \\
$c$ (\AA) & 10.336(2)  &19.537(7)  \\
$\beta$ (deg) & 104.88(2) & 90 \\
Volume (\AA$^3$) & 284.24(14) &  267.99(11)\\

$Z$ &  2 & 4 \\
Wavelength (\AA) & 0.40255 & 0.40255 \\
Detector distance (mm) & 118.37 & 118.31 \\
No. of images & 543 & 543 \\
$(\sin\theta/\lambda)_{max}$ (\AA$^{-1}$) & 0.812432 & 0.823225 \\

Absorption, $\mu$ (mm$^{-1}$) & 6.673 & 6.965 \\
$T_{min}$, $T_{max}$ & 0.225, 0.235 & 0.476, 1 \\

Criterion of observability & $I > 3\sigma(I)$ & $I > 3\sigma(I)$ \\

No. of reflections measured &  1902 & 784 \\
No. of unique reflections (obs/all) & 319/416 & 94/124 \\

$R_{int}$ & 0.0612 & 0.0724 \\

No. of parameters & 123& 14 \\

$R_F$ (obs/all) & 0.0676/0.0676 & 0.0562/0.0562 \\
$wR_F$ (all) & 0.1687 & 0.1518 \\

GoF (obs/all) & 3.4317/3.4317 & 4.4203/4.4203 \\

$\Delta \rho_{min}$, $\Delta \rho_{max}$ (e \AA$^{-3}$)
& -2.48, 2.50
& -2.48, 1.81 \\

\hline\hline
\end{tabular}
\end{table*}

\bibliography{Biblio}